\documentclass[usenatbib]{mn2e}
\usepackage{epsfig}
\usepackage{times}
\title[3D dust and gas radiative transfer simulations] {The dusty {\sc mocassin}: fully self-consistent 3D photoionisation and dust radiative transfer models}\author[Ercolano et al.]{B. Ercolano$^1$, M. J. Barlow$^1$, P. J. Storey$^1$\\
$^1$Department of Physics and Astronomy, University College London, Gower Street, London WC1E~6BT, UK\\}
\date{Received:}

\begin{document}
\maketitle

%%%%%%%%%%%%%%%%%%%%%%%%%%%%%%%%%%%%%%%%%%%%%%%%%%%%%%%%%%%%%%%%%%%%%%%%%%%
\begin{abstract}
We present the first 3D Monte Carlo (MC) photoionisation code to include a fully self-consistent treatment of dust radiative transfer (RT) within a photoionised region. This is the latest development (Version~2.0) of the gas-only photoionisation code {\sc mocassin} (Ercolano et al., 2003a), and employs a stochastic approach to the transport of radiation, allowing both the primary and secondary components of the radiation field to be treated self-consistently, whilst accounting for the scattering of radiation by dust grains mixed with the gas, as well as the absorption and emission of radiation by both the gas and the dust components. An escape probability method is implemented for the transfer of resonance lines that may be absorbed by the grains, thus contributing to their energy balance. The energetics of the co-existing dust and gas components must also take into account the effects of dust-gas collisions and photoelectric emission from the dust grains, which are dependent on the grain charge. These are included in our code using the average grain potential approximation scheme. 

A set of rigorous benchmark tests have been carried out for dust-only spherically symmetric geometries and 2D disk configurations. {\sc mocassin}'s results are found to be in agreement with those obtained by well established dust-only RT codes that employ various approaches to the solution of the RT problem. 

A model of the dust and of the photoionised gas components of the planetary nebula (PN) NGC~3918 is also presented as a means of testing the correct functioning of the RT procedures in a case where both gas and dust opacities are present. The two components are coupled via the heating of dust grains by the absorption of both UV continuum photons and resonance line photons emitted by the gas. The {\sc mocassin} results show agreement with those of a 1D dust and gas model of this nebula published previously, showing the reliability of the new code, which can be applied to a variety of astrophysical environments. 

\end{abstract}

\nokeywords

%%%%%%%%%%%%%%%%%%%%%%%%%%%%%%%%%%%%%%%%%%%%%%%%%%%%%%%%%%%%%%%%%%%%%%%%%%%

\section{Introduction}

The presence of dust grains in ionised plasma environments can have significant effects on the radiative transfer (RT) and influence the physical conditions of the gas, as the grains compete with the gas for the absorption of continuum UV photons.  As well as being heated via the absorption of such photons, the dust grains are also heated by nebular resonance line radiation emitted by the gas. Moreover, photo-electric emission from dust grains and dust-grain collisions provide additional cooling and heating channels for both the gas and dust components. The evident coupling between the co-existing dust and gas phases can only be treated properly in a photoionisation code by the incorporation, in the RT, of scattering, absorption and emission of radiation by dust particles mixed with the gas. 

Whilst some 1D photoionisation codes, such as {\sc cloudy} (Ferland et al. 1998, Van Hoof et al. 2004) and the Harrington code (Harrington et al., 1988, hereafter H88), include a treatment for the RT of dust, and a number of 2D or 3D pure-dust RT codes using MC or ray tracing techniques are available (e.g. Bjorkman \& Wood, 2001, or Pascucci et al., 2004 for a recent review), to our knowledge, no existing 3D photoionisation code self-consistently treats the transfer of both the stellar and diffuse components of the radiation field through a region where dust particles are mixed with the gas. 

We present here a new version (2.0) of the 3D photoionisation code {\sc mocassin} (Ercolano et al., 2003a, hereafter Paper~{\sc i}), designed to achieve the above by employing a stochastic approach to simultaneously solve the dust and gas RT. As well as calculating emission line spectra and the 3D ionisation and electron temperature and density structure of a nebula, the code can also determine accurate dust temperatures and spectral energy distributions (SEDs) by treating discrete grain sizes and different grain species separately, compared to the frequently used approximation of having a single grain as representative of an ensemble of grains.

The new code is described here together with its solutions for a number of pure-dust benchmark problems and the pseudo-benchmark case of a dust and gas model for the PN~NGC~3918. Section~2 contains a description of the MC techniques applied to the solution of the RT problem in a dusty photoionised region. Dust-only benchmark models and their solutions are presented in Section~3 for the 1D and the 2D cases. A 3D model of the photoionised region and the thermal IR emission of the PN~NGC~3918 is presented in Section~4, together with a comparison of our results with those obtained by previous 1D modelling and the available observational data. A final summary is presented in Section~5.

\section{Dust radiative transfer in the photoionised region}
\label{sub:drt}

Paper~{\sc i} presented the fully 3D photoionisation code, {\sc mocassin}, which employs a MC approach to the transfer of radiation. The description of the radiation field as composed of a discrete number of monochromatic packets of energy $E_0$ (Abbot \& Lucy, 1985) allows for both primary and secondary components of the radiation field to be determined exactly. The mean intensity of the radiation field, $J_{\nu}$, is derived at each location using the MC estimator constructed by Lucy (1999) as given by equation~11 of Paper~{\sc i}. The path of an energy packet through the grid completely defines its contribution to $J_{\nu}$ at each location; this reduces the RT problem to the computation of the energy packets' trajectories as they diffuse out of the nebula. In the case of a pure-gas nebula, only absorption and re-emission of the packets by the ions in the gas need be considered, and therefore the local gas opacities and emissivities, which depend on the electron temperatures and ionic fractions, are the necessary ingredients of the computation. 

However, dust grains that are mixed with the gas in the ionised region provide an extra source of opacity and, as well as competing for the absorption of UV continuum radiation, are also capable of absorbing line photons emitted by resonant transitions. In the remainder of this section the discussion will be limited to the transfer of the continuum radiation, with the description of the resonant line transfer escape probability method being given in Section~\ref{sec:rlt}. 

The random walk of stellar energy packets injected into the nebula is characterised by absorption and re-emission events due to the atoms and ions in the gas, as well as absorption, re-emission and scattering events due to the dust grains. The locations of the interactions are calculated using the (second) MC method described in Section~2.4 of Paper~{\sc i}, with the random path of a packet between events, $l$, being described by
\begin{equation}
\tau_{\nu_P} = (\kappa_{\nu_P}^{gas} + \kappa_{\nu_P}^{ext})\cdot{l}
\end{equation}
where $\tau_{\nu_{\rm P}}$ is the total optical depth of the gas and dust to the next interaction at frequency $\nu_{\rm P}$ along the packet's direction of travel, $\kappa_{\nu_{\rm P}}^{gas}$ is the local gas opacity at $\nu_{\rm P}$, while $\kappa_{\nu}^{ext}$~=~$\kappa_{\nu}^{sca}$+$\kappa_{\nu}^{abs}$ is the dust extinction opacity at $\nu_{\rm P}$ due to scattering and absorption by the dust grains.

This first MC choice can only tell us whether a packet is going to interact at a given location or is going to continue into the next grid cell. The nature of the interaction is however not known at this stage. The probability of a packet of frequency $\nu_{\rm P}$ undergoing a scattering event, $P_{sca}(\nu_{\rm P})$, is given by the ratio of the scattering to the total opacity, comprising both the gas and dust extinction opacities:
\begin{equation}
P_{sca}(\nu_{\rm P}) = \frac{\kappa_{\nu_{\rm P}}^{sca}}{\kappa_{\nu_{\rm P}}^{gas}+\kappa_{\nu_{\rm P}}^{ext}}
\label{eq:pdust}
\end{equation}
If the second MC test, performed at this stage, results in the packet being absorbed, then the process continues as for the gas-only version of the code described in Paper~{\sc i}.  

Energy conservation is enforced by ensuring that the packets that are absorbed by the dust grain are re-emitted in the new direction immediately after the absorption event. The frequency of the re-emitted packet is selected according to the local gas+dust emissivities, with the frequency distribution of the re-emitted packets determined from the normalised cumulative probability density function. A grain of size $a$ and species $s$ will provide a contribution to the local continuous emission at frequency $\nu$ equal to
\begin{equation}
j_{\rm a,s}^{\rm dust}(\nu) = 4\pi\,a^2\,Q_{\rm a,s}^{\rm abs}(\nu)\,B_{\nu}(T_{\rm a,s})
\end{equation}
where $Q_{\rm a,s}^{\rm abs}(\nu)$ is the grain absorption efficiency at frequency $\nu$ and $B_{\nu}(T_{\rm a,s})$ is the Planck function at the grain temperature. If $j^{\rm gas}(\nu)$ is the frequency-dependent gas continuum emissivity, corrected for the contribution due to He~{\sc i} and He~{\sc ii} Lyman lines, as described in Section~2.6 of Paper~{\sc i}, then the total emission from a volume element at frequency $\nu$ is given by 
\begin{equation}
\eta(\nu) = j^{\rm gas}(\nu)\cdot N_{\rm gas} + \sum_{\rm s}\int_{\rm a}j_{\rm a,s}^{\rm dust}n_{\rm a,s}da
\end{equation}
where $N_{\rm gas}$ is the gas number density, $n_{\rm a,s}$ is the number density of grains of species $s$ and size $a$, while the summation and the integration in the second term of the RHS are over the grain species included and the size distribution, respectively. 

The normalised cumulative probability density function, $p_{\nu}$, is then calculated as follows
\begin{equation}
p_{\nu} = \frac{\int_{\nu_{\rm min}}^{\nu} \eta(\nu') d\nu'}{\int_{\nu_{\rm min}}^{\nu_{\rm max}} \eta(\nu') d\nu'}
\end{equation}
where the integral in the denominator is over the full spectral range considered and defined by $\nu_{\rm min}$ and $\nu_{\rm max}$.

A packet undergoing a scattering event will simply be re-directed, with its frequency remaining unchanged. Unless a phase function is specified, scattering is assumed to be isotropic, with the new random directions being selected according to equation~4 in Paper~{\sc i}. In the non-isotropic case the Henyey-Greenstein (HG) phase function (Henyey \& Greenstein, 1941) can be used in its standard or modified form (Henney \& Axon, 1995) to determine a packet's scattering angle. In this case the anisotropy parameter $g$ must be specified by the user.

A packet is transferred through the grid until no more interactions with gas or dust are possible. In general, emission lines are assumed to escape without further interaction; special attention is given to resonance lines, which can be absorbed by the dust grains, as we discuss later in Section~\ref{sec:rlt}. 

\subsection{The dust opacities}

{\sc mocassin} uses standard Mie scattering theory (e.g. Kattawar \& Plass 1967; Wickramasinghe 1972; Andriesse 1979) to evaluate the effective absorption and scattering efficiencies, $Q_{abs}(a, \lambda)$ and $Q_{sca}(a, \lambda)$, for a grain of radius $a$ at wavelength $\lambda$, from the dielectric constants of any given material. The optical constants library distributed with the source code includes the most frequently used astronomical dust species, such as warm and cold silicates (Ossenkopf, Henning \& Mathis 1992), astronomical silicates and graphite (Draine \& Laor 1993, Draine \& Lee 1984, Hageman, Gudat \& Kunz 1974), amorphous carbon (Hanner 1988), $\alpha$SiC (P\'egouri\'e 1988) and another seventeen species. However, supplementary/alternative data can be easily supplied by the user. 

With the exception of silicates, graphite and SiC  by Draine \& Laor (1993) and graphite by Hageman, Gudat \& Kunz (1974), the data provided does not extend to the far UV (FUV) domain. At ionising energies, such data must be supplemented in order to allow for the absorption of starlight that is the main heating process for the grains, which in turn may provide a significant source of opacity to the nebula.

\subsection{Grain sizes and Chemistry}

Grains of different sizes and species are treated separately, with individual grain temperatures being determined for each grain radius and species included. It has been previously demonstrated that approximating a grain mixture with a representative composite grain can yield misleading conclusions (e.g. Van Hoof et al. 2004), given that the emission due to the smaller, hotter grains may be hard to reproduce accurately using this technique. 

There is no limit to the number of sizes that may be included in a given model, except perhaps that intrinsically imposed by the available computing resources. Any type of grain size distribution and chemical mixture can be handled by the code. Furthermore, these do not need to be defined homogeneously across the grid, but multiple chemistry and/or grain-size distributions can be specified at different locations. The latter option was implemented in view of the dual circumstellar dust chemistry (oxygen- and carbon-rich) found in some PNe ionised by Wolf-Rayet central stars (e.g. Cohen et al., 2002; De Marco et al. 2004). 

\subsection{Determination of dust temperatures}
\label{sec:gt}

The local grain temperatures for each species and grain size are determined by the equation of radiative equilibrium between all cooling and heating channels. The grains are mainly cooled via their temperature-dependent thermal emission and heated by the absorption of UV photons from the radiation field, as well as by resonance line photons. As will be described in more detail in the next section, resonance line transfer is treated using an escape probability method similar to that described by Cohen, Harrington \& Hess (1984) and later modified and applied by H88 to their dust model for the PN~NGC~3918. At each location in the grid, we estimate $f_{\rm e}$, the position-dependent fraction of photons in a given resonance line that escape, and, if $G_{\rm L}$ is the rate of energy generation in line L, then, for any given dust species, the heating contribution from line L to a grain of radius $a$ is given by 
\begin{equation}
H_L = G_L\cdot(1-f_{\rm e})\cdot Q_{abs}(a,\lambda_L)/\kappa_{\rm d}(\lambda_L)
\label{eq:hl}
\end{equation}
where $\lambda_{\rm L}$ is the central wavelength of the resonance line and $\kappa_{\rm d}(\lambda_L)$ is the mean dust opacity of that species, which determines the intensity within the line. 

The continuous part of the mean intensity of the radiation field, $J(\lambda)$, is estimated as discussed in Section~\ref{sub:drt}; using the above and neglecting photoelectric emission processes and gas-dust collisions (see Section~\ref{sub:gd}), the radiative equilibrium equation can be written as follows:
\begin{eqnarray}
\int_{\nu_{min}}^{\nu_{max}}{B_{\lambda}(T)\,Q_{abs}(a,\lambda)\,d\lambda} = ~~~~~~~~~~~~~~~~~~~~~~~~~~~~~~~~~~~~~~~~~~~~~\\ \nonumber
\int_{\nu_{min}}^{\nu_{max}}{J_{\lambda}(T)\,Q_{abs}(a,\lambda)\,d\lambda} + \sum_L{H_L}
 \end{eqnarray}   
where $B_{\lambda}(T)$ is the usual Planck function evaluated at the grain temperature and integrated over the frequency range delimited by $\nu_min$ and $\nu_max$ and the summation in the second term of the right-hand-side of the equation is over all the resonance lines included in the model. 

\subsubsection{Resonant line transfer}
\label{sec:rlt}

In the previous gas-only version of the {\sc mocassin} code, line packets could escape immediately after they had been created. The only exception was made in the case of He~{\sc i} Lyman lines, that can further ionise H$^0$, and He~{\sc ii} Lyman lines that can ionise both H$^0$ and He$^0$. In the presence of dust grains, however, care must also be taken in the transfer of resonance emission lines such as, for example, H~{\sc i}~Ly$\alpha$, C~{\sc iv}~$\lambda$1549, N~{\sc v}~$\lambda$1240, C~{\sc ii}~$\lambda$1336, Si~{\sc iv}~$\lambda$1397 and Mg~{\sc ii}~$\lambda$2800. These lines, amongst others, can undergo many resonant scattering events, resulting in a much longer random walk before they can reach the edge of the nebula and escape, with a consequently higher probability of being absorbed by dust. 

The emergent resonant line intensities will, therefore, be attenuated by dust absorption, with the absorbed fraction contributing to the heating of the grains. This is an extra complication arising from the interaction between the gas and the dust phase, it is nevertheless essential to carefully treat this process, which can often have sizable effects on the emergent continuous spectral energy distribution by raising the grain temperatures, as well as affecting the observed intensities of the emergent resonance emission lines.

H88 employed a generalisation to 1D spherical geometry of the plane-parallel escape probability approach of Cohen et al. (1984), whereby the resonance line photons are assumed either to be absorbed locally by the dust or be scattered into the line wings and escape in a single flight. We have further generalised the argument to arbitrary geometry and implemented it within the 3D {\sc mocassin} code, for the transfer of some important resonance emission lines, including Ly$\alpha$~1216, C~{\sc ii}~1336, C~{\sc iv}~1549, N~{\sc v}~1240, Mg~{\sc ii}~2800 and Si~{\sc iv}~1397. Any other line of interest can easily be added to this list by providing the relevant atomic data to the code's library.

The basic idea consists of obtaining the fraction of escaping resonance line photons, $f_{\rm e}$, at each grid location for each resonance line. This quantity can then be used in equation~\ref{eq:hl} above in order to estimate the contribution of a given resonance line to the dust heating. The contribution from this grid location to the dust-attenuated nebular luminosity in this resonance line can subsequently be obtained by multiplying this factor by the value calculated using the formal solution or the full MC method, as described in Paper~{\sc i}. 

For each resonance line $P_e[\tau(\hat{u})]$, the probability of escape without further interaction of a photon scattered in direction $\hat{u}$, is first calculated. Values of $P_e[\tau(\hat{u})]$ are calculated at each location in the grid, in a number of radial directions emerging from the centre of the cell\footnote{64 directions were used at each cell for the model presented in Section~\ref{sec:3918}.} using the following expression (Kwan \& Krolik, 1981):
\begin{eqnarray}
P_e[\tau(\hat{u})] = \frac{1}{\tau\sqrt{\pi}(1.2+b)}&~~~~~~&\tau\geq{1}\\ \nonumber
P_e[\tau(\hat{u})] = \frac{1-e^{2\tau}}{2\tau}     &~~~~~~&\tau{<}{1}
\end{eqnarray}
where $b\,=\,\sqrt{log(\tau)}/(1+\tau/\tau_{\rm w})$, $\tau$ = $\tau(\hat{u})$ is the optical depth at line centre in direction $\hat{u}$ and $\tau_w$~=~10$^5$.  Assuming the scattering to be isotropic, we obtain $P_{\rm e}$~=~$<P_{\rm e}[\tau(\hat{u})]>$, the escape probability averaged over all directions. Using this value of $P_{\rm e}$ and equation A~8 of Cohen et al. (1984) we can estimate the total fraction of escaping resonance line photons, $f_{\rm e}$. It is perhaps worth noting at this point that our approach differs somewhat from that of H88, who used equation~A1 of Cohen et al. (1984) to calculate $M_2[\tau(\hat{u})]$, the dust-free escape probability for a resonance line photon travelling in direction $\hat{u}$, and then obtained a direction averaged escape probability $P_e$, by taking the mean of the $M_2[\tau(\hat{u})]$ values for 40 radial directions.

It should, finally, be stated that in a dust-only simulation the only source of grain heating comes from the continuous radiation field, so that in this case the above becomes irrelevant and no overheads are introduced in the execution time/memory requirements.

We compared the treatment described above with the exact results of Hummer \& Kunasz (1980) by constructing a plane parallel model in {\sc mocassin} to mimic a uniform slab of optical thickness $T$. Table~\ref{tab:reslinecomp} shows the results obtained by {\sc mocassin} and those given in Table~2 of Hummer \& Kunasz (1980) for the fraction of escaping resonance line photons from mid-plane. We included the cases when $T\,=\,10^4$ and $10^6$, and for a range of $\beta$ values, where $\beta$ is the ratio of continuum to line opacity, $\kappa_d$/$\kappa_l$. For the $T~=~10^4$ case we also include the results given in Table~4 of Cohen et al. (1984). All our values are within 20\% of the exact values, apart from the $T~=~10^6$, $\beta=4.03\times10^{-7}$ case, where a discrepancy of 34\% is obtained. The results of Hummer \& Kunasz (1980) depend also on the parameter $a$, which is the ratio of natural to doppler line widths. Only the latter is taken into account by our treatment and that of Cohen et al. (1984), therefore our results are best compared with those listed by Hummer \& Kunasz (1980) for the lowest value of $a$ (4.7$\times$10$^{-4}$). We note that although we have reasonable agreement with the results of Hummer \& Kunasz (1980) for the $a=4.7\times 10^{-4}$ and $T\leq 10^6$ case, the neglect of the damping wings for transitions with large optical depth means that we will necessarily overestimate the fraction of photons escaping for larger values of $a$ and $T$.

\begin{table}
\caption{Resonance line photon escaping fractions from a slab geometry of optical thickness $T$\,=\,10$^4$ and 10$^6$, calculated for a range of continuum- to line-opacity ratios, $\beta$\,=\,$\kappa_d$/$\kappa_l$. {\sc mocassin}'s results for the mid-plane values ({\sc moc}) are compared to those obtained by Cohen et al. (1984) (CHH84, for $T$\,=\,10$^4$) and to the exact treatment of Hummer \& Kunasz, 1980 (HK80).}
\label{tab:reslinecomp}
\begin{center}
\setlength\tabcolsep{5pt}
\begin{tabular}{cccccccc}
\hline\noalign{\smallskip}
\multicolumn{4}{c}{$T~=~10^4$} & \multicolumn{4}{c}{$T~=~10^6$} \\
$\beta$ & {\sc moc} & CHH84 & HK80& $\beta$ & {\sc moc} & CHH84 & HK80  \\
\hline
 1.00-6 & 0.851 & 0.836 & 0.908 & 3.16-8 & 0.774 & -- & 0.804 \\
 3.16-6 & 0.644 & 0.606 & 0.750 & 6.70-8 & 0.616 & -- & 0.645 \\
 5.00-6 & 0.533 & 0.557 & 0.650 & 1.00-7 & 0.519 & -- & 0.544 \\
 1.00-5 & 0.364 & 0.386 & 0.436 & 1.34-7 & 0.445 & -- & 0.451 \\
 2.00-5 & 0.222 & 0.239 & 0.279 & 2.69-7 & 0.285 & -- & 0.252 \\
 5.00-5 & 0.103 & 0.112 & 0.097 & 4.03-7 & 0.211 & -- & 0.158 \\
\hline
\end{tabular}
\end{center}
\end{table}

\subsubsection{Additional heating and cooling channels}
\label{sub:gd}

A number of additional microphysical processes come into play when dust and gas interactions are considered. Photoelectric emission from the surface of dust grains can provide an additional heating channel for the photoionised gas in a variety of astrophysical environments, including H~{\sc ii} regions and PNe (e.g. Maciel \& Pottasch 1982, Baldwin et al. 1991, Borkowski \& Harrington 1991, Ercolano et al., 2003c). When this process is included, the energy carried out by the photoelectrons must also be accounted for in the thermal balance equations of the grains as a heat sink. The calculation of photoelectric thresholds and yields requires a knowledge of the individual grain potentials (Weingartner \& Draine 2001). Furthermore, energy is also exchanged between the gas and the grains via collisions that will cool the gas and heat the grains in the case when the latter have the lower temperature. 

The processes listed above are treated by {\sc mocassin } using the average grain potential approximation, as described by Baldwin et al. (1991), that is based on the detailed balance between grain charge gain and loss rates. Whilst this is an excellent approximation for large grains, it may fail to provide an accurate description for very small grains (e.g. Weingartner and Draine, 2001). 

As photoelectric emission from dust grains and gas-grain collisional processes are irrelevant to the applications presented in this work, we refer to a forthcoming paper (Ercolano et al., in preparation) for a more detailed description of their implementation in the {\sc mocassin} code.

Finally, we note that at present grain charge transfer is not included in our code. This may provide a net recombination process for the heavy elements (e.g. Draine \& Sutin 1987) in the H$^+$ region.

\subsection{Emerging Spectral Energy Distribution}

The emergent SED is obtained by collecting the energy packets as they reach the outer edge of the grid and escape. Frequency and direction information is retained for each packet, so that it is later possible to account for orientation effects, that are often of vital importance for asymmetric models. 

As mentioned above, the spectral distribution of the radiation re-processed by the grains is calculated by integrating equation~3 over the size distribution and abundances of the grain mixture, with $B_{\nu}(T_{a,s})$ being evaluated for each grain of species $s$ and radius $a$. This means that the emerging continuum energy packets will automatically provide the correct spectral shape for the chosen dust chemistry and discrete size distribution chosen in the model. 

This is certainly an improvement over the practice of calculating SEDs using a single dust temperature to represent an ensemble of grains of various species and sizes. Whilst this may have been a necessity in the past, due to limited computing power, it may result in misleading conclusions, as the emission from hotter, smaller grains may not be accurately reproduced (e.g. Van Hoof et al. 2004).

Furthermore, very small grains that have a small heat capacity can undergo large temperature fluctuations, also known as {\it temperature spikes}. This process, first suggested by Greenberg (1968), is included in the current version of {\sc mocassin}, using the method of Guhathakurta \& Draine (1989) in the implementation of Sylvester et al. (1997). This effect is generally important for grains with typical radii of approximately 10~{\AA}, however the minimum grain sizes for which the process should be taken into account is treated as an input parameter in order to provide the user with some control over the overheads introduced. 

\section{Pure Dust Benchmarks}

{\sc mocassin}'s dust and gas processes are fully integrated, it is however possible to run the code for models containing only one or the other types of process. The pure gas photoionisation version was benchmarked against established 1D photoionisation codes according to the Lexington-Meudon test models (Paper~{\sc i}), and we have checked that, in the absence of dust, these can still be correctly reproduced by the current version. Furthermore, the new dust RT procedures were thoroughly tested by running dust-only simulations for a set of 1D and 2D benchmark problems, as described below. 

As for previous versions, {\sc mocassin} Version~2.0 is also fully parallel and can run efficiently on multi-processors or Beowulf systems. Dust-only models, however, converge much more quickly than full photionisation ones, thanks the to dust opacities being virtually independent of temperature. The amount of memory required to run such models is also much less, which makes {\sc mocassin} also suitable for running on smaller single-processor machines. As a guideline, all benchmark models presented in this section were computed on a  Pentium~M 1.70~GHz portable PC with 1GB of RAM. Models converged to acceptable accuracy within 1~hr or less, with longer computational times sometimes being required to smooth out numerical errors that are intrinsic to the MC technique employed. 

\subsection{Spherically symmetric benchmarks}

\begin{table}
\caption{Input parameters for spherically symmetric benchmark models. For all models, the ratio of the outer radius $R_{\rm out}$ to the inner radius of the shell Y~=~1000; $N_{\rm dust}$ is the number density of the grain; $L_*$ and $T_{\rm eff}$ are the luminosity and the effective temperature of the ionising source, which radiates as a blackbody; $a_0$ is the radius of the dust grains.}
\label{tab:input1d}
\begin{center}
\setlength\tabcolsep{5pt}
\begin{tabular}{cccccc}
\hline\noalign{\smallskip}
$\tau_{\rm 1 {\mu}m}$ &  $R_{\rm out}$        & $N_{\rm dust}$ &  $L_*$        & $T_{\rm eff}$ & $a_0$    \\
                      & [cm]                  & [cm$^{-3}$]    & [L$_{\odot}$] & [K]           &[${\mu}m$]\\
\hline
    1.0               & 2.18$\times$10$^{17}$ &  1.12$\times$10$^{-8}$ & 1.00$\times$10$^{4}$ & 2500   &  0.16    \\
    10.0              & 2.20$\times$10$^{17}$ &  1.20$\times$10$^{-7}$ & 1.00$\times$10$^{4}$ & 2500   &  0.16    \\
    100.0             & 2.35$\times$10$^{17}$ &  1.04$\times$10$^{-6}$ & 1.00$\times$10$^{4}$ & 2500   &  0.16    \\
\hline
\end{tabular}
\end{center}
\end{table}

\begin{figure}
\begin{center}
\begin{minipage}[t]{8.5cm}
\includegraphics[width=8.5cm]{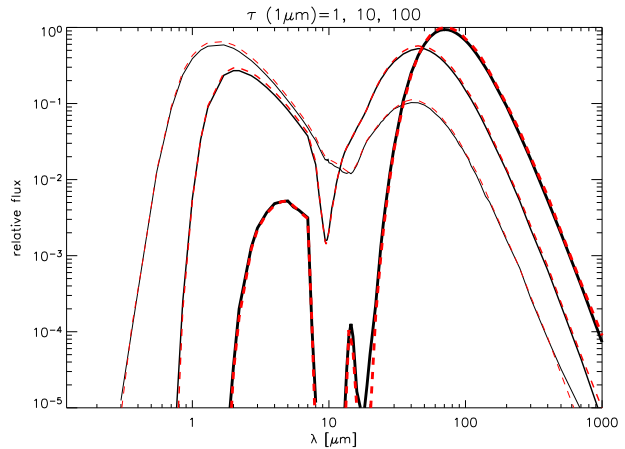}
\end{minipage}
\caption[]{Spherically symmetric, homogeneous benchmark models for a 10$^4$\,L$_{\odot}$ star surrounded by dust shells with $\tau_1$=1, 10 and 100. The thicker lines have been used to represent the more optically thick models. The SEDs predicted by {\sc dusty} (dashed lines) are compared to those obtained by {\sc mocassin} (solid line).}
\label{fig:dusty}
\end{center}
\end{figure}

A set of spherically symmetric benchmarks problems have been provided by Ivezic et al. (1997), where solutions obtained by three RT codes, including the widely used {\sc dusty} (Ivezic \& Elitzur, 1997), are compared. The codes, each implementing a different numerical scheme, were found to agree to better than 0.1 per cent. The radial dust temperature distributions and the emerging SEDs for a number of these benchmark tests were calculated using {\sc mocassin} and compared to results obtained with the current version of  {\sc dusty} (2.01). Some preliminary benchmark results have already been presented by Ercolano et al. (2005). 

We present here three spherically symmetric homogeneous dust shell models, illuminated by a central blackbody source and containing silicate grains of radius 0.16$\mu$m. Optical constants for the `astronomical silicates' of Draine \& Lee (1984) were used in the {\sc mocassin} models, as these are also included in {\sc dusty}'s standard optical constants library. The scattering was assumed to be isotropic for all benchmark models. The optical thickness of the shell at $\lambda$~=~1~$\mu$m for the three models was $\tau_1$=1, 10 and 100. The input parameters are summarised in Table~\ref{tab:input1d}.

Figure~\ref{fig:dusty} shows a comparison between the SEDs predicted by {\sc mocassin} (solid lines) and those predicted by {\sc dusty} (Ivezic et al., 1997; dashed line) for the $\tau_{1}$~=~1 (thin lines), $\tau_{1}$~=~10 (medium line) and $\tau_{1}$~=~100 (thick line) models. The emerging SEDs are plotted, by analogy with Figure~2 of Ivezic et al. (1997), as dimensionless, distance- and luminosity-independent spectral shapes $\lambda\,F_{\lambda}$/$\int{F_{\lambda}d\lambda}$. Taking into account that a completely different approach to the radiative transfer and the calculation of the SEDs is used by the two codes, the results obtained from the benchmarking are in very good agreement. The radial distributions of grain temperatures (not shown here) were also found to be in excellent agreement. 

\subsection{2D disk benchmarks}

\begin{figure*}
\begin{center}
\label{fig:2dsed}
\begin{minipage}[t]{8.5cm}
\includegraphics[width=8.5cm]{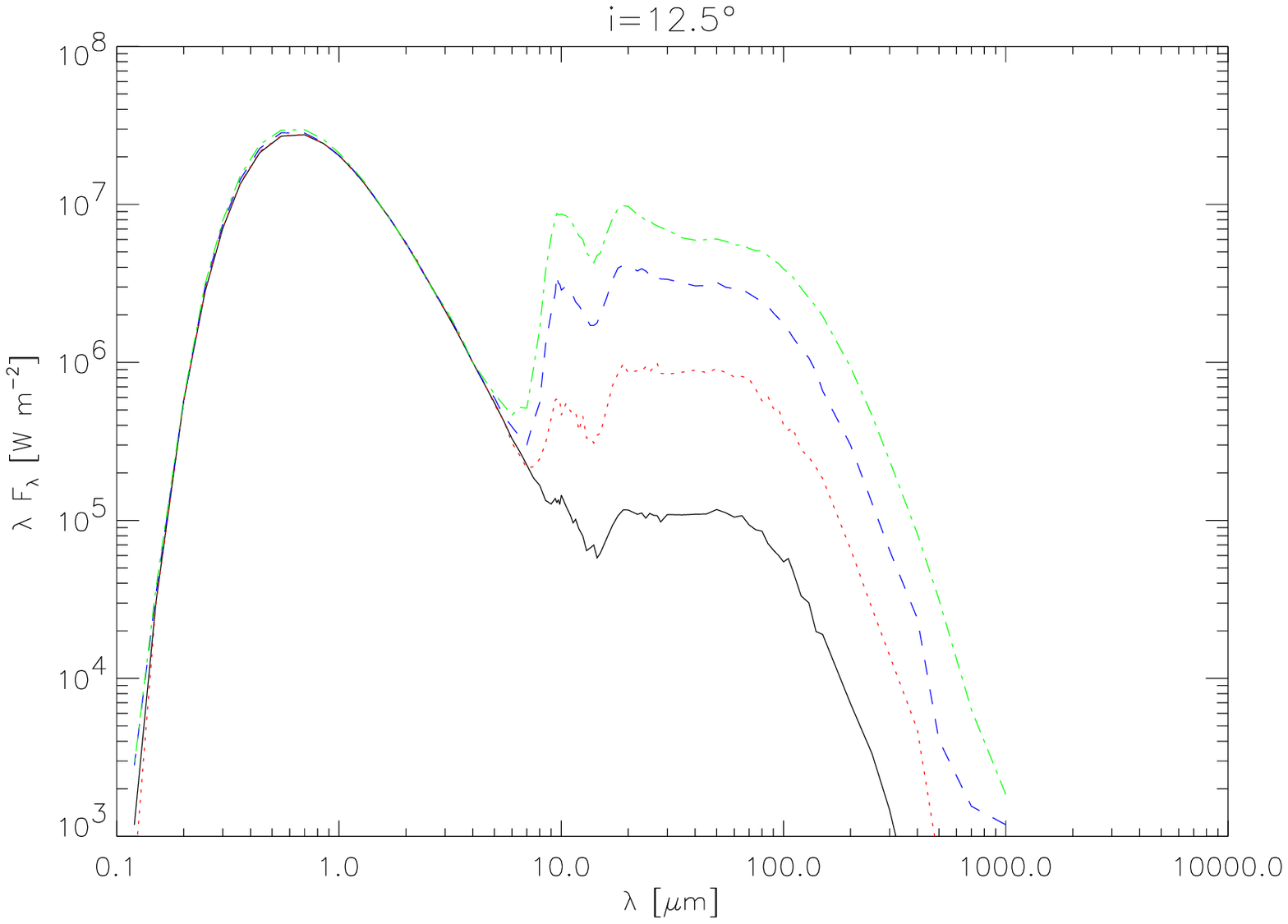}
\end{minipage}
\begin{minipage}[t]{8.5cm}
\includegraphics[width=8.5cm]{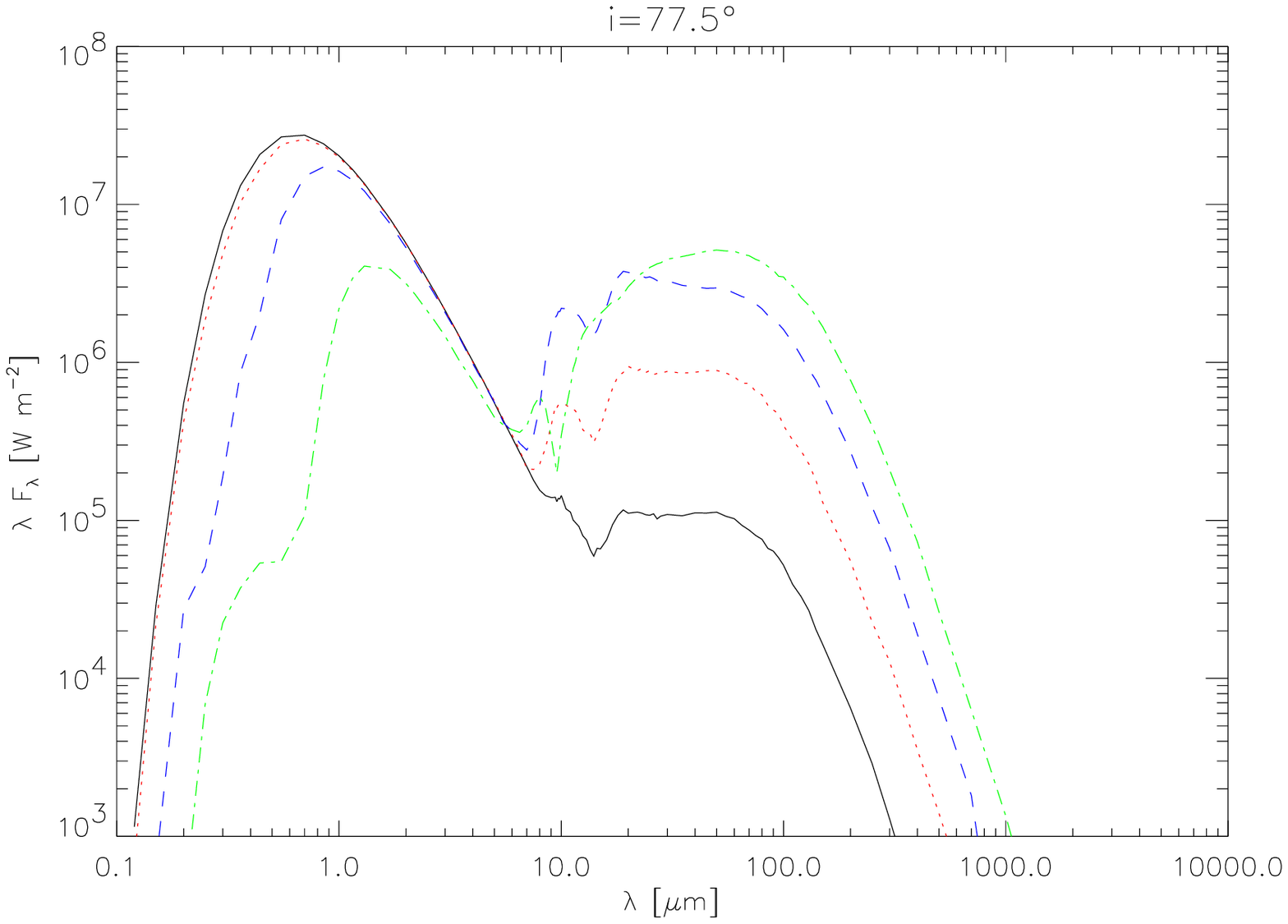}
\end{minipage}
\caption[]{Spectral energy distributions for disk inclinations i~=~12.5$^{o}$ (left panel) and i~=~77.5$^{o}$ (right panel) for the $\tau_{\rm V}$~=~0.1 (solid line), $\tau_{\rm V}$~=~1 (dotted line), $\tau_{\rm V}$~=~10 (dashed line) and  $\tau_{\rm V}$~=~100 (dot-dash line)  benchmark models around a 5800~K, 1~L$_{\odot}$ . This figure is directly comparable to figure~7 of Pascucci et al. (2004).}
\end{center}
\end{figure*}

\begin{table}
\caption{Input parameters for 2D disk benchmark models. $\tau_{\rm V}$: mid-plane visual optical depth ($\lambda$\,=\,550\,nm);  $R_{\rm in}$: inner disk radius; $R_{\rm out}$: outer disk radius; $z_{\rm d}$: disk thickness; $L_*[L_{\odot}]$: central star luminosity; $T_{\rm eff}$[K]: central star effective temperature;  $a_0[{\mu}$m$]$: grain radius. The density distribution was calculated using the parameters listed and equation~4 of Pascucci et al. (2004).}
\label{tab:2dbenchin}
\begin{center}
\setlength\tabcolsep{5pt}
\begin{tabular}{ccccccc}
\hline\noalign{\smallskip}
$\tau_{V}$ &$R_{\rm in}$ & $R_{\rm out}$ & $z_d$ &  $L_*$        & $T_{\rm eff}$ & $a_0$  \\
           & $[$AU$]$    &$[$AU$]$ & $[$AU$]$ & $[L_{\odot}]$    & $[$K$]$        & $[{\mu}m]$ \\
\hline
    0.1     & 1 & 1000          & 125      & 1.02    & 5800          &  0.12     \\
    1.0     & 1 & 1000          & 125      & 1.02    & 5800          &  0.12        \\
    10.0    & 1 & 1000          & 125      & 1.02    & 5800          &  0.12       \\
    100.0   & 1 & 1000          & 125      & 1.02    & 5800          &  0.12        \\
\hline
\end{tabular}
\end{center}
\label{Tab1a}
\end{table}

Four benchmark solutions for the continuum RT in a 2D disk configuration were presented by Pascucci et al. (2004). They tested the behaviour of five different codes for a set of well-defined 2D models having different visual ($\lambda$=550nm) optical depths, $\tau_{\rm V}$, including a high optical depth ($\tau_{\rm V}$=100) and strong (isotropic) scattering case.  All models consisted of a point-source embedded in a circumstellar dust disk with an inner dust-free cavity. The dust is composed of spherical astronomical silicate grains (Draine \& Lee, 1984) with a radius of 0.12$\mu$m and density of 3.6\,g\,cm$^{-3}$. Results from various codes for the radius-dependent grain temperatures and the emergent SEDs for a number of radial directions and disk inclinations were compared by Pascucci et al. (2004) and found to be in reasonable agreement for most test cases. All benchmark tests were also successfully reproduced by the {\sc mocassin} code as briefly shown in this section. The disk geometry employed is typical of those generally used for the study of disks around T Tauri stars (Natta et al. 2000) and described by equation~4 of Pascucci et al. (2004). A summary of all input parameters is given in Table~\ref{tab:2dbenchin}, with the symbols used being defined in the caption.

%The radial dust temperature distributions calculated by {\sc mocassin} for the most optically thick case are shown in figure~\ref{fig:2dt} for an angle $\theta$~=~2.5$^{o}$, close to the disk mid-plane. This figure is directly comparable with figure~5 of Pascucci et al. (2004) and demonstrates the good agreement of {\sc mocassin}'s solution with that obtained by the other codes. 

Figure~\ref{fig:2dsed} shows the SEDs ($\lambda\,F_{\lambda}$ $[$W m$^{-2}]$ against $\lambda$ [$\mu$m]) for two disk inclinations: one almost face-on (i~=~12.5$^{o}$ -- left panel), and the other almost edge-on (i~=~77.5$^{o}$ -- right panel --) for the $\tau_{\rm V}$\,=\,0.1 (solid line), 1 (dotted line), 10 (dashed-line) and 100 (dash-dot line) benchmark models around a 5800~K 1~L$_{\odot}$ star. This figure is directly comparable with figure~7 of Pascucci et al. 2004 and shows good agreement for all cases. 

The published benchmark solutions for the radial dust temperature distributions were also successfully matched by {\sc mocassin}. 

\section{Modelling the thermal emission of the PN NGC 3918}
\label{sec:3918}
The benchmark models described above were designed to test the solution of the dust RT problem in the absence of gas. A set of benchmarks to test the correct RT solution of dust and gas mixed together is however not available. In this work we will present a fully-self consistent 3D photoionisation model of the PN NGC 3918, where dust grains are mixed with the gas inside the ionised region. The geometry of this model and the detailed dust properties are based on the work of Clegg et al. (1987) and H88, who used a volume weighted combination of two spherically symmetric dust and gas density distributions fed into a 1D photoionisation code that could also account for dust absorption. Their results are treated here as a `realistic' benchmark case that can be compared with {\sc mocassin}'s solution, within the limits imposed by the different approaches employed by the two codes and the different atomic data sets available to them.

\subsection{Photoionisation models of NGC 3918}

A pure-gas photoionisation model of NGC 3918 was originally presented by Clegg et al. (1987), where the nebula was approximated by two dense (optically thick) cones immersed in a sphere of more diffuse (optically thin) gas. This biconical model, which was successful in reproducing most of the spectroscopic constraints available for this object, became unrealistic once the morphology of NGC~3918 had been disclosed by later observations (e.g. Corradi et al., 1999). Nevertheless, the same density distribution was used by Ercolano et al. (2003b) to construct a 3D model of NGC~3918 using {\sc mocassin}; the model was used as a 'realistic' 3D benchmark for the then new photoionisation code. Furthermore, differences between the two sets of results for the ionisation structure and emission line spectra were identified in that paper and discussed in the light of the effects of the radiation field being transferred self-consistently through the discontinuity between the cones and the surrounding material. A new model for the pure-gas photoionised region of NGC~3918 was also presented by Ercolano et al (2003b). This provided a better fit to the available spatio-kinematical constraints, as well as reproducing most of the observed spectroscopic characteristics. However, in the present work, NGC~3918 is treated merely as a benchmark model to test the coupling of dust and gas in {\sc mocassin}'s radiative transfer and, for this purpose, only the original bi-conical model needs to be considered. 

\subsection{The dust model}

\begin{table}
\caption{ Input parameters for the {\sc mocassin} dust and gas model of NGC~3918. All gas abundances are given by number. See text for a description of the symbols used.}
\label{tab:ngc3918in}
\centering
\begin{tabular}{llll}
\hline
\noalign{\smallskip}
L$_{\rm *}$   & 7224 L$_{\odot}$ & N/H  & 1.5$\times$10$^{-4}$ \\ 
T$_{\rm eff}$ &  140000 K        & O/H  & 5.0$\times$10$^{-4}$ \\ 
$\rho$        & 2.2 g\,cm$^{-3}$ & Ne/H & 1.2$\times$10$^{-4}$ \\
a$_{min}$     & 0.04$\mu$m       & Mg/H & 1.4$\times$10$^{-5}$ \\
a$_{max}$     & 0.40$\mu$m       & Si/H & 1.0$\times$10$^{-5}$ \\
R             & 0.0005           & S/H  & 1.6$\times$10$^{-5}$ \\ 
He/H & 0.107                     & Ar/H & 2.0$\times$10$^{-6}$ \\
C/H  & 8.0$\times$10$^{-4}$      & Fe/H & 3.7$\times$10$^{-7}$ \\
\noalign{\smallskip}
\hline
\end{tabular}
\end{table}

The thermal IR emission of NGC~3918 was modelled by H88, who added dust grains to the 1D photoionisation model described above. They explored a number of size distributions and dust chemistries and agreed on a best-fitting model, which they named the 'primary model', consisting of spherical graphite grains, with $\rho=2.2\,$g\,cm$^{-3}$ (Draine \& Lee, 1984 and Hagemann, Gudat \& Kunz, 1974) with the standard MRN dust parameters of $a_{\rm min}$=0.04$\mu$m and  $a_{\rm max}$=0.40$\mu$m and a dust-to-hydrogen ratio, $R$, of 0.0005 by mass. 

The same input parameters as for the pure-gas photoionisation model were used for the gas and dust model, except for the luminosity of the central star, which was increased by 4.7\% in order to compensate for the stellar radiation absorbed by the dust (H88 argued that the same result could have also been obtained by decreasing the gas density by 2.3\%). Due to the total optical depth of the dust being less than unity in this model, the effects of the inclusion of the dust opacity on the emission line spectrum, ionisation and gas thermal structure of the nebula are small. Results for all the gas properties were found to be similar to those obtained in the dust-free models, as described by Ercolano et al. (2003b). 

The central star parameters and the elemental abundances used for the dust and gas 3D {\sc mocassin} model are summarised in Table~\ref{tab:ngc3918in}, the parameters controlling the shape of bicone were given in Table~1 of Ercolano et al. (2003b). 

It is worth noting at this point that the effects of the scattering opacity on the radiation field were not taken into account by H88. For ease of comparison, this process was also neglected in the initial {\sc mocassin} models, to be reactivated in the subsequent models, in order to investigate its effects on the final results. We found that for the relatively low dust optical depths of NGC~3918, dust scattering had only a small effect on the final dust temperatures and SEDs. 

Photoelectric emission from dust grains and gas-grain collisions were not included in the models of H88 and, therefore, were also neglected in our models. Finally, as the MRN size distribution was truncated to a lower limit of 0.04$\mu$m, the grains are not expected to be affected by temperature spiking (see section~2.4). 

\subsection{Comparison with previous models and available IR observations}

The observational IR data available for NGC~3918, as described by H88, include four sets in the wavelength range 8 to 100~${\mu}$m. Broadband photometry at 12, 25, 60 and 100$\mu$m is available from the {\it IRAS Point Source Catalog} (1985), and at 36 and 70 $\mu$m from Moseley (1980). Cohen \& Barlow (1980) gave narrow band photometry from 8$\mu$m to 20$\mu$m, while a similar spectral range (8$\mu$m to 23$\mu$m) was also covered with the {\it IRAS} Low-Resolution Spectrometer (Pottasch et al., 1986). We refer to H88 and to the respective references listed above for further description of these data sets. 

\subsubsection{The emergent spectral energy distribution}

The SEDs from our model were convolved with the IRAS broadband filter transmission functions and compared with the line- and colour-corrected {\it IRAS Point Source Catalog} fluxes. A full description of the line and colour corrections applied is given by H88. Following H88, the Cohen \& Barlow (1980) flux at 9.6$\mu$m is to be preferred to the {\it IRAS} 12$\mu$m flux, given the large ionic emission line contribution to the latter. Table~\ref{tab:comparesed} lists {\sc mocassin}'s model predictions in the second row and the observed fluxes in the first row, H88's model predictions (row~3) are also given for comparison.

Although the IR flux distribution is slightly different from H88 for the 5 wavelength regions, the total IR ($\sim$5-300$\mu$m) luminosities are similar for the two codes (within 5\%). The remaining differences are however acceptable in light of the different approach emplyed by our code with regards to the dust model, where we treat the individual grain sizes independently in the RT, as well in the final temperature and SED calculations, and also with regards to the RT between the two optical depth regions that is treated self-consistently in our 3D simulation. 

\begin{table}
\caption{ IR fluxes [Jy] predicted by the {\sc mocassin} model and by H88, compared with line- and coloured-corrected {\it IRAS Point Source Catalog} fluxes and the 9.6$\mu$m flux measurement of Cohen \& Barlow (1980) }
\label{tab:comparesed}
\centering
\begin{tabular}{lccccc}
\hline
\noalign{\smallskip}
                                   &   9.6$\mu$m   &   12$\mu$m   &   25$\mu$m   &   60$\mu$m   &   100$\mu$m   \\
\hline
 Observed                          &   1.04        &   4.7        &   38.4       &   45.6       &   16.5        \\
{\sc mocassin} model               &   3.03        &   4.1        &   29.0       &   57.0       &   13.3        \\
H88 model                          &   0.95        &   2.63       &   40.0       &   47.1       &   11.0        \\
\noalign{\smallskip}
\hline
\end{tabular}
\end{table}

\subsubsection{Dust attenuation of UV resonance lines}
The emergent resonance emission line intensities are heavily attenuated by dust absorption. The escape probability method adopted for the RT of the resonance lines has already been described in Section~\ref{sec:rlt}. Table~\ref{tab:darl} lists {\sc mocassin}'s predictions for the dust-attenuated resonance line intensities and percentage destruction fractions in each of the two sectors. The observed values and the line intensities and percentage destruction fractions predicted by H88 for the two sectors are also listed for comparison. All line intensities are given relative to H$\beta$, on a scale where I(H$\beta$)~=~100. 

\begin{table}
\caption{ Dust attenuation of resonance line radiation. The line flux intensities, I($\lambda$) are given relative to H$\beta$ on a scale where I(H$\beta$)\,=\,100. }
\label{tab:darl}
\centering
\begin{tabular}{lccccccc}
\hline
\noalign{\smallskip}
                                   & \multicolumn{4}{c}{Percent Destroyed}  & \multicolumn{3}{c}{I($\lambda$) (attenuated)} \\                            
\hline      
                                   & \multicolumn{2}{c}{\sc moc.}    &\multicolumn{2}{c}{ H88}          &{\sc moc.} & H88  & Obs \\
                                   &   Thick       &   Thin         &  Thick       &   Thin       &               &                 \\
\hline
Ly$\alpha$ 1216                    &    68         &    44          & 80           & 23           &   859         & 728 & *         \\  
C~{\sc iv} 1550                    &    54         &    38          & 45           & 21           &   1025        & 950 & 512       \\  
N~{\sc v} 1240                     &    39         &    32          & 25           & 20           &   24          & 52  & 46        \\  
Mg~{\sc ii} 2800                   &    48         &    18          & 32           & -            &   42          & 29  & *         \\  
Si~{\sc iv} 1400                   &    65         &    30          & 40           & 13           &   9.1         & 5.7 & 9         \\  
\noalign{\smallskip}
\hline
\end{tabular}
\end{table}

There are evident differences between our results and those of H88. These are not surprising and are due to a variety of factors. The atomic data set available to {\sc mocassin} has naturally been expanded in the 17 years since the publication of H88's results, The only meaningful comparison between the two codes can be made in the case of Ly$\alpha$, here we see a reasonable agreement (to within 12\%) in the predicted total attenuated flux for this line. Even for this line there are differences in the predicted destruction fractions, with {\sc mocassin} predicting a lower fraction in the thick sector and a higher fraction in the thin sector. This is easily understood by considering that in our 3D model resonance line packets emitted in the thin sector can cross over to the thicker one, with a higher probability of being absorbed by the dust in the thick sector, and vice-versa for lines emitted in the thick sector. 

The infrared energy balance is summarised in Table~\ref{tab:enbudget} which lists the major contributions to dust heating. In agreement with H88's conclusions we also found that Ly$\alpha$ is the dominant source of heating in the optically thick region, followed by UV continuous radiation. In the thin sector, where the lower gas denisties naturally result in lower line emission and where the destruction factors are much reduced, Ly$\alpha$ provides about 24\% of the total grain heating, with the continuous radiation becoming the dominant source here. Heating by absorption of the other resonance line photons is also significant in both sectors. The small dicrepancies between our results and H88's can once again be explained in terms of the 3D transfer as well as differences in the atomic data sets. 

\begin{table}
\caption{ Percentage dust heating contributions. }
\label{tab:heatingbudgets}
\centering
\begin{tabular}{lcccccc}
\hline
\noalign{\smallskip}
Energy source                      & \multicolumn{6}{c}{Percentage Heating Contribution}  \\                            
\hline      
                                   & \multicolumn{2}{c}{Thick}       &    \multicolumn{2}{c}{Thin}     &  \multicolumn{2}{c}{Total} \\
                                   & {\sc moc.}    &    H88          & {\sc moc.}  &    H88            & {\sc moc.}    &    H88     \\
\hline
Ly$\alpha$ 1216                    &   55.7        &   67.6          &   24.2      &   17.8            &   47.1     &   57.3      \\  
UV continuum                       &   15.9        &   17.2          &   39.4      &   58.5            &   22.5     &   25.8      \\
C~{\sc iv} 1550                    &   26.0        &   15.0          &   35.3      &   21.8            &   28.6     &   16.4      \\
Other lines                        &   2.4         &   0.2           &    1.1      &   1.9             &   1.8      &   0.5       \\
\noalign{\smallskip}
\hline
\end{tabular}
\end{table}

\subsubsection{The dust temperatures}
\label{sec:dt}

\begin{figure}
\begin{center}
\begin{minipage}[t]{8.5cm}
\includegraphics[width=8.5cm]{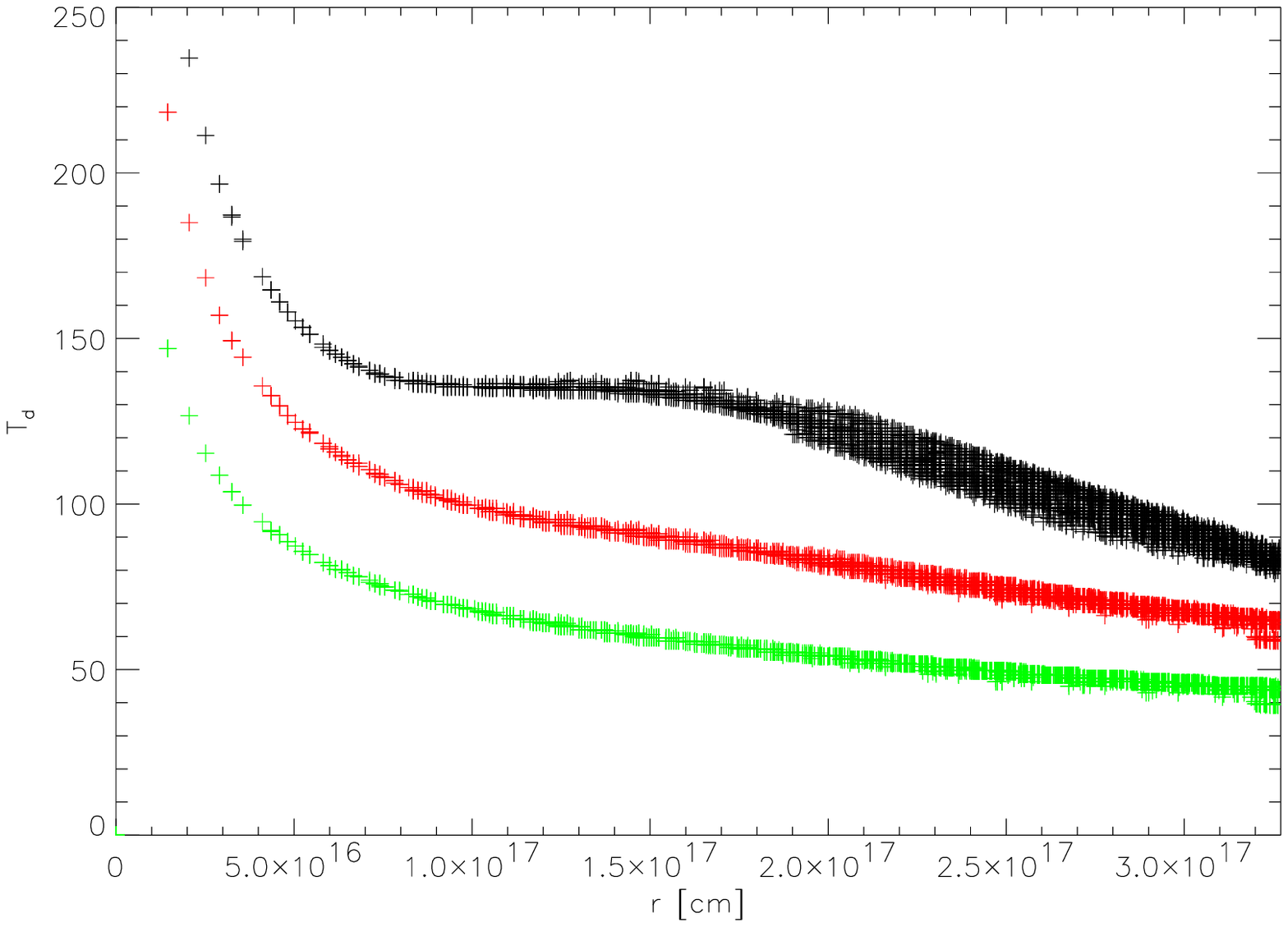}
\end{minipage}
\begin{minipage}[t]{8.5cm}
\includegraphics[width=8.5cm]{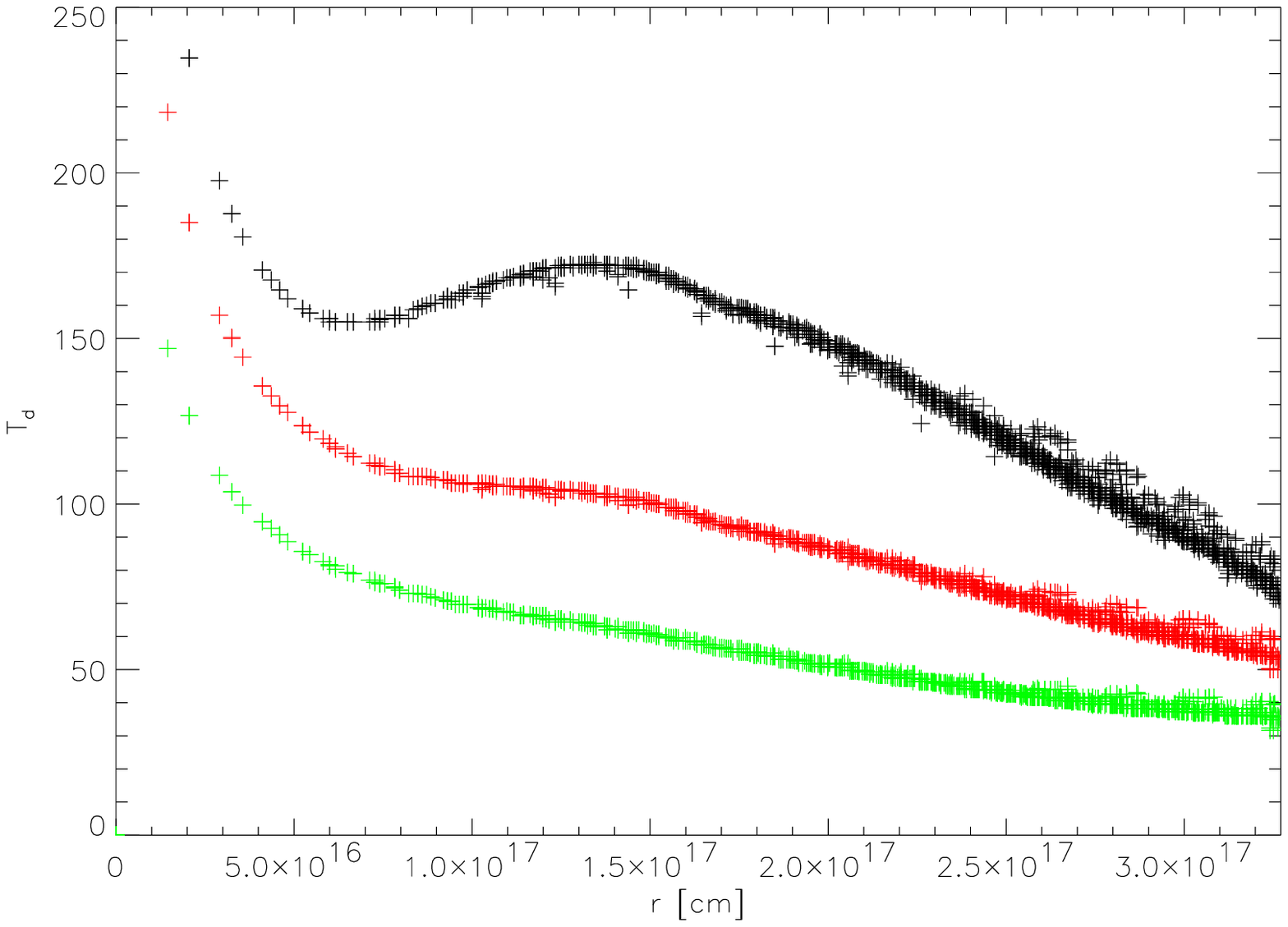}
\end{minipage}
\caption[]{Radial grain temperatures for grains of radius 0.04$\mu$m (hottest), 0.1$\mu$m (medium) and 0.40$\mu$m (coolest). The cells in thin sectors are plotted in the top panel and those from the thick sector are in the bottom panel. This figure is comparable to figure~3 of H88. }
\label{fig:gt}
\end{center}
\end{figure}

Grain temperatures for each grain size are determined from the equation of radiative equilibrium, as discussed in Section~\ref{sec:gt}. Figure~\ref{fig:gt} shows the radius-dependent temperature distributions for grains of radius 0.04$\mu$m (hottest), 0.1$\mu$m and 0.40$\mu$m (coolest). Each data point represents the temperature of a grain in a given cell and all cells belonging to the optically thin density phase are plotted in the top panel, whilst those belonging to the optically thick bicone are plotted in the bottom panel. The difference in the grain temperatures in cells belonging to the two different optical depth regions is evident, with the more opaque regions being naturally hotter. The difference is further accentuated by the extra heating provided by the resonance line photons, a process that is more efficient in the optically thick sector than in the thin one. The scatter of the data is only partially due to numerical noise; it is also a result of radiation transfer between the optically thick cones and the surrounding thinner regions. 

While the grain temperatures in the two sectors seem to be in agreement with those shown in figure~3 of H88, a closer inspection reveals that our temperatures in the thick sector are consistently lower, whilst those in the thin sector are consistently higher than predicted by H88. The very small discrepancies are due to the fact that in the case of H88's 1D composite models the resonance line emission produced in one sector will only be able to heat the grains within the same sector. In our 3D model, which is self-consistent at the boundaries between the two sectors, packets can be transfered throughout the volume and, therefore, dust grains in the optically thin region are also heated by resonance line packets produced in the thick biconical regions. Naturally, the converse is also true, but the density imbalance between the two sectors practically results in dust-heating {\it leaking} from the optically thick to the surrounding optically thin region. All our model results are consistent with this interpretation.

\section {Conclusions}

We have presented a new tool for the analysis of dusty photoionised plasma. The dust-enhanced version of the fully 3D photoionisation code {\sc mocassin} is suitable for the interpretation of observational data from such environments, including emission line spectra, thermal SEDs and narrow or broad-band images at optical or IR wavelengths.

Our results demonstrate the reliability of the new version of {\sc mocassin} that employs a MC approach to solving the RT problem for dust and gas simultaneously. The benchmark tests showed that {\sc mocassin} was able to reproduce {\sc dusty}'s solutions for grain temperatures and SEDs for a number of spherically symmetric pure-dust benchmark models (Ivezic et al., 1997). The code also achieved a good level of accuracy in matching the results of a number of well established 2D and 3D dust RT codes for a set of 2D disk benchmarks proposed by Pascucci et al. (2004). 

A fully self-consistent 3D photoionisation and dust model was constructed for the PN~NGC~3918, which was treated as a realistic dust and gas benchmark case, thanks to the existence of a previous 1D dust and gas photoionisation model by H88. Our results were shown to be in reasonable agreement with those obtained by H88, within the limits imposed by the fundamental differences between the two approaches and the latter's need to adopt a composite model of two volume weighted 1D models to describe a 3D structure. 

The new version of {\sc mocassin} (Version~2.0) is now fully functional and it will be made freely available to the astronomical community shortly after the publication of this article\footnote{Further information can be obtained from the authors via email: be@star.ucl.ac.uk}.

\vspace{4mm}
\noindent
{\bf Acknowledgements}
This work was partly carried out on the Enigma supercomputer at the HiPerSPACE Computing Centre, UCL, which is funded by the UK Particle Physics and Astronomy Research Council. 
We thank Dr J. Yates for helpful discussion during the 1D benchmarking phase of the pure-dust models. We thank the anonymous referee for helpful comments. 

\bibliographystyle{mn2e}

\bibliography{references}

\bsp
\end{document}